# Single-Crystal AlN Wafer-Based Bulk Acoustic Resonators for Piezoelectric Power Conversion

Ziqian Yao†, Clarissa Daniel‡, Kaicheng Pan§, Tzu-Hsuan Hsu†, Heather Chang‡,
Mark S Goorsky§, Juan Rivas-Davila‡, and Ruochen Lu†
†The University of Texas at Austin, US, ‡Stanford University, US, §The University of California Los Angeles, US
hanson.yao@utexas.edu

*Summary*— In this work, we demonstrate the first single-crystal aluminum nitride (AlN) wafer-based thickness-extensional (TE) mode bulk acoustic resonator for piezoelectric power conversion. The device exhibits a high series resonance 3-dB quality factor ($Q$) of 1677 and an electromechanical coupling coefficient ($k^2$) of 6.1%, highlighting the strong potential of AlN resonators for efficient power conversion. To suppress in-band spurious modes, a grounded ring structure is proposed and experimentally validated. The measured frequency-domain impedance response shows a spurious suppression of the spectrum above the resonance at 13.52 MHz. A comparative analysis with prior PZT, LN, and LT-based resonators indicates that AlN achieves a competitive figure of merit and $f·Q$ product, while its material thermal conductivity is orders of magnitude higher than that of the incumbent piezoelectric power-converter resonators. The power-handling capability is expected to be superior in AlN single-crystal wafers and will be demonstrated in ongoing experiments. These results suggest that AlN offers a promising platform for compact, robust piezoelectric power converters and next-generation power electronic systems.

## I. Introduction

Piezoelectric power conversion has recently emerged as a promising approach for replacing bulky inductors in power converters by exploiting miniature acoustic resonators. Drawing on design principles established in radio-frequency acoustic devices, this strategy enables compact form factors, improved efficiency, and power density [1-4].

The state-of-the-art (SoA) piezoelectric materials for power conversion (Table 1) are evaluated using key performance metrics, including the electromechanical coupling coefficient ($k^2$), quality factor ($Q$), and thermal stability. Early power resonators relied on lead zirconate titanate (PZT) to achieve high figures of merit (FoM = $k^2Q$), with ($k^2$) reaching 30-40%; however, their performance is often limited by nonlinearity and poor thermal robustness [5-9]. Lithium niobate (LN) later demonstrated advantages in FoM, linearity, and power handling, enabling compact high-power converters in the multi-kW regime [10-14]. Nevertheless, LN's large temperature coefficient of frequency (TCF ≈ −70 ppm/K) can cause self-heating induced frequency drift during sustained operation, often necessitating external heat sinks or active cooling [15], which makes it less suitable for applications demanding high thermal stability. More recent work on lithium tantalate (LT) has shown improved thermal stability due to its lower TCF [16];

| Material | Mode | e | $\varepsilon_r$ | FoM [$e^2/(\varepsilon·c)$] | Loss Tangent | Thermal Conductivity (W/m·K) | TCF (ppm/K) | Wafer Available |
|---|---|---|---|---|---|---|---|---|
| PZT-5A | LE | 5.35 | 826 | 0.33% | 0.018 | 1.3 | −35 | Y |
| $Sc_{0.3}Al_{0.7}N$ | LE | 0.70 | 21.1 | 0.84% | 0.0036 | 2.5 | −35 | N |
| $Sc_{0.3}Al_{0.7}N$ | TE | 2.38 | 21.1 | 9.7% | 0.0036 | 2.5 | −35 | N |
| LN | LE | 1.83 | 38.6 | 4.8% | 0.00013 | 2.24 | -70 | Y |
| LN | TE | 4.53 | 38.6 | 31.2% | 0.00013 | 2.68 | -69 | Y |
| LT | TE | 3.25 | 41.7 | 10.7% | 0.00013 | 3.8 | -12 | Y |
| AlN | LE | 0.59 | 10.3 | 0.96% | 0.001 | 316 | −28 | Y |
| **AlN** | **TE** | **1.47** | **10.3** | **6.0%** | **0.001** | **316** | **−28** | **Y** |

**Table 1.** Comparison between AlN and other acoustic platforms for piezoelectric power converters.

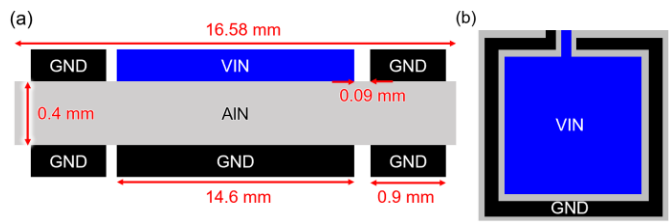

**Fig. 1.** Schematics and key dimensions of (a) side-view of AlN thickness extensional (TE) mode bulk acoustic resonators with (b) top-down view.

however, its relatively low thermal conductivity can still lead to heat accumulation at elevated power levels.

In this work, we address thermal accumulation in piezoelectric power resonators by utilizing single-crystal aluminum nitride (AlN). AlN offers intrinsically high thermal conductivity, typically on the order of hundreds of W·m⁻¹·K⁻¹ for high-quality single crystals, which enables efficient heat dissipation, reduces self-heating, and improves frequency stability under sustained high-power operation [17-20]. In contrast, piezoelectric materials such LN, LT, sputtered ScAlN, and PZT exhibit much lower thermal conductivity, generally in the single-digit W·m⁻¹·K⁻¹ range depending on composition and crystal orientation, which makes them more prone to localized temperature rise and thermally driven drift at comparable power density [21-25].

The substantially higher thermal conductivity of single-crystal AlN originates from its wurtzite crystal structure, strong bonding, and light constituent atoms. These features support high phonon group velocities and long phonon mean free paths in high-purity crystals [26-27]. By comparison, LN and LT exhibit stronger anharmonic phonon scattering and correspondingly lower lattice thermal conductivity [28-29],

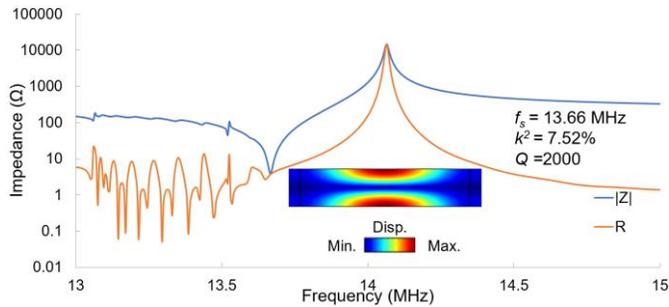

**Fig. 2.** Finite element simulated frequency domain impedance and resistance, along with inset figure showing the cross-sectional displacement at while PZT and related perovskite ferroelectrics further experience additional phonon scattering associated with structural complexity and disorder, including defect and domain-related scattering mechanisms [30-33]. Motivated by these thermal-transport advantages and the availability of single-crystal AlN wafers, we demonstrate the first prototype resonators targeted for thermally robust piezoelectric power conversion at low-MHz operation.

## II. Design and Simulation

Fig. 1 illustrates the schematic of the proposed AlN resonator. The device employs a pair of rectangular electrodes patterned on the top and bottom surfaces of a 0.4-mm-thick AlN wafer. On the top side, the electrode consists of a 14.6-mm-wide central VIN region surrounded by a 0.9-mm-wide grounded ring, separated by a 0.09-mm ring gap. The bottom electrode mirrors the top electrode layout, with segmented grounded regions aligned to the top-side pattern. This grounded-ring configuration is adapted from a previous thickness-extensional (TE) mode resonator design reported in [34-36], which implemented a grounded ring structure on LN to suppress in-band spurious modes.

To validate the design, we performed finite element analysis (FEA) in COMSOL to simulate the electrical admittance and mechanical displacement mode shapes (Fig. 2). A constant $Q$ of 2000 was incorporated as mechanical damping in the model. From the simulated mode shape, we identify the fundamental $S_1$ mode at the series resonance frequency $f_s = 13.66$ MHz. The simulated $S_1$ mode exhibits dominant out-of-plane thickness motion with the peak displacement concentrated near the device center, consistent with efficient TE-mode excitation. No strong in-band parasitic modes are observed in the simulated impedance and resistance response, indicating that the grounded-ring architecture effectively mitigates spurious mode coupling in the target frequency range. Based on the simulated series and parallel resonance frequencies, we extract $k^2 = 7.52\%$.

## III. Device Fabrication

After validating the resonator design in COMSOL, the AlN resonators were fabricated and integrated using a standard cleanroom process flow (Fig. 3) on a 2-inch, 0.4-mm-thick, (0001)-oriented single-crystal AlN wafer from HexaTech. The wafer exhibits high crystalline quality, with an HRXRD rocking-curve width below 100 arcsec. The substrate was first solvent-cleaned in acetone and isopropyl alcohol (IPA). Photoresist was then spin-coated and patterned by photolithography to define the top-side electrodes, followed by electron-beam evaporation of the metal stack. Gold (Au) was selected for its high electrical conductivity, and a 15-nm chromium (Cr) adhesion layer was deposited prior to the 400-nm Au layer. After lift-off, the wafer was flipped for backside processing. The cleaning, lithography, and metallization steps were repeated with backside alignment to define the grounded ring features, ensuring reliable grounding while minimizing unintended electrical feedthrough between isolated structures. The completed devices were then separated into individual dies using an automated wafer dicing saw.

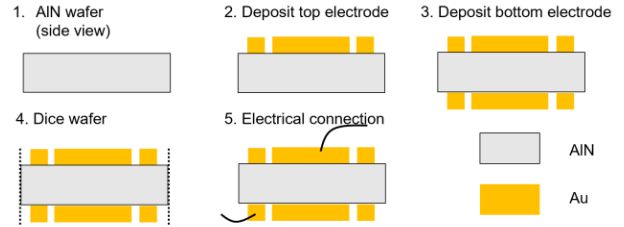

**Fig. 3.** Fabrication and integration process flow chart

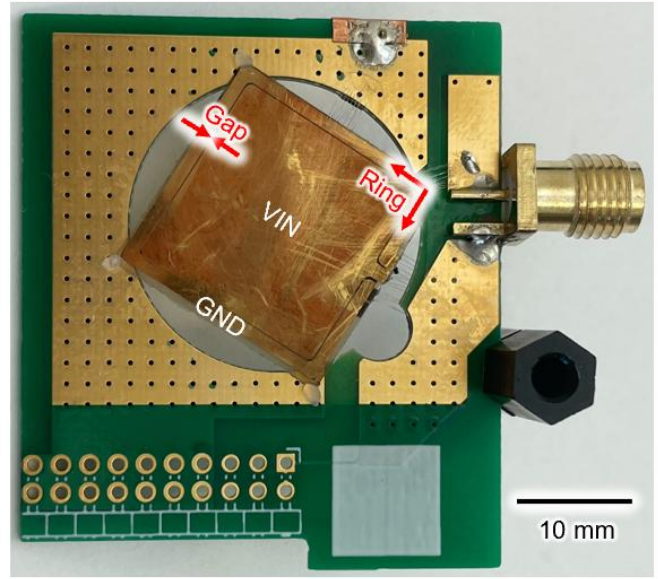

**Fig. 4.** Fabricated resonators integrated for power electronic testing.

For electrical characterization and packaging (Fig. 4), each die was mounted onto an FR-4 printed circuit board (PCB) using UV-cure epoxy. This mounting approach provides mechanical damping and establishes electrical grounding through the bottom pads, while keeping the central active electrode suspended in air to enable efficient TE-mode excitation and reduce acoustic energy leakage.

## IV. Measurement and Discussion

The integrated device is characterized using a vector network analyzer (VNA) through the SMA port on the FR-4 PCB board. The measured frequency-domain impedance magnitude and resistance are presented in Fig. 5, showing a clear resonance at 13.52 MHz with an extracted $k^2 = 6.1\%$. Compared to the idealized finite-element simulations, additional in-band spurious features are observed. We attribute these spurious modes primarily to non-ideal boundary conditions and surface-induced scattering, which are not captured in simulations that assume perfectly smooth and symmetric surfaces. In particular, the current devices are fabricated on single-side-polished AlN wafers, which can introduce asymmetric surface roughness between the top and bottom interfaces. Such asymmetry can perturb the intended thickness-extensional displacement distribution, increase acoustic scattering, and enable unintended coupling to parasitic modes that remain weak in an ideal, perfectly symmetric structure.

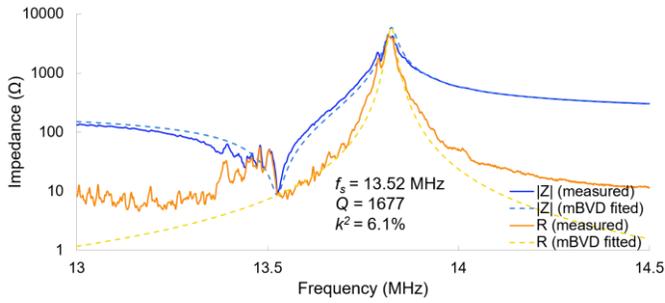

**Fig. 5.** Measured frequency domain impedance and resistance of fabricated AlN resonator, featuring spurious-free spectrum above the resonance at 13.52 MHz, high $Q$ of 1677 and $k^2$ of 6.1%.

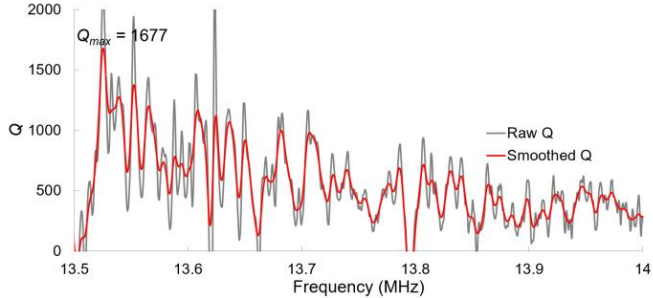

**Fig. 6.** Raw and smoothed Bode $Q$ of AlN resonator.

| Platform | Ref. | $f_s$ (MHz) | $k^2$ | Q | TCF (ppm/K) | FoM (Q×$k^2$) | $f_s$×Q (×$10^{10}$) |
|---|---|---|---|---|---|---|---|
| PZT-TE | Boles et al | 0.61 | 31% | 2500 | N/A | 775 | 0.15 |
| PZT-Radial | Daniel et al | 0.174 | 42.5% | 1198 | -35 | 740 | 0.021 |
| LN-TE | Touhami et al | 6.28 | 25.5% | 3700 | N/A | 944 | 2.32 |
| LN-TS | Nguyen et al | 5.94 | 45% | 3500 | N/A | 1575 | 2.08 |
| P3F LN-TE | Yao et al | 19.23 | 29% | 3187 | -70.3 | 928 | 6.13 |
| LT-TE | Yao et al | 6.5 | 8.8% | 1968 | -13.56 | 173 | 1.28 |
| AlN | This Work | 13.52 | 6.1% | 1677 | -28 | 103 | 2.27 |

**Table 2.** Comparison of this work to prior art. This work features good FoM and $f \cdot Q$ as resonators, promising for power conversion applications.

In addition, device packaging and electrical parasitics can contribute to deviations from simulation. The PCB mount, epoxy layer, and wire/solder interfaces introduce additional mechanical loading and electrical feedthrough paths that can slightly shift the resonance frequency and change the impedance. Despite these non-idealities, the measured response remains largely spurious-free around the target resonance, indicating that the grounded-ring architecture effectively suppresses strong in-band parasitic resonances and stabilizes the electrical boundary conditions at the device periphery.

Fig. 6 shows the Bode $Q$ plot of the measured AlN resonator extracted from the phase derivative of the measured S-parameters. After smoothing, the device exhibits a stable and high-quality factor, with a peak smoothed 3-dB $Q$ of 1677 at resonance. This measured $Q$ is lower than the constant $Q$ assumption used in simulation, which is expected because the experimental result includes additional loss mechanisms from electrode resistance, imperfect boundary conditions, and energy leakage into the package. Nevertheless, the measured $f \cdot Q$ product remains competitive for low-MHz TE-mode operation, and the achieved $k^2$ and $Q$ together yield a favorable figure of merit for piezoelectric power conversion.

The performance of this work is compared with representative PZT- and LN-based resonators in Table 2. While AlN exhibits a lower $k^2$ than typical PZT- and LN-based platforms, it achieves a competitive $f \cdot Q$ product. More importantly, AlN provides a materials advantage for power conversion through its substantially higher thermal conductivity, which enables more efficient heat spreading across the substrate during sustained high-power operation. This characteristic is expected to mitigate local hot spots, reduce self-heating-induced frequency drift, and improve long-term stability compared to lower-thermal-conductivity piezoelectric platforms.

Future work will focus on wafer surface quality, device geometry, and packaging optimization. Temperature testing and nonlinear power-handling will be analyzed prior to power converter integration. Adopting double-side-polished AlN wafers is expected to reduce surface asymmetry and minimize scattering-induced parasitic resonances. In parallel, electrode and layout optimization will be explored to further reduce electrical feedthrough and improve mode purity. Packaging will also be improved by better controlling epoxy thickness and reducing variability in mechanical damping, allowing intrinsic resonator performance to be more clearly separated from mount-dependent effects. To directly evaluate power-conversion suitability, ongoing measurements will include thermal resilience and nonlinear power-handling characterization. Thermal cycling over a wide temperature range will quantify frequency drift, $Q$ stability, and any degradation of electrode metal or interface defects under repeated thermal stress. Large-signal testing will be used to assess nonlinear behavior at high drive levels, including resonance distortion, changes in effective $Q$, and power-dependent frequency shift. Finally, breakdown-related limits will be characterized by measuring the maximum sustainable voltage and current at resonance, providing a practical assessment of device reliability and power capability under conditions representative of piezoelectric power conversion.

## V. Conclusions

This work demonstrates an AlN TE-mode bulk acoustic resonator for piezoelectric power conversion. The proposed device achieves a resonance frequency of 13.52 MHz, a high $Q$ of 1677, and a $k^2$ of 6.1%. A grounded-ring electrode structure is employed to mitigate in-band spurious modes, enabling a competitive $f \cdot Q$ product and FoM for power-conversion applications. In addition to the demonstrated resonance performance, AlN provides a key materials advantage due to its intrinsically high thermal conductivity, which is expected to enhance heat spreading during sustained operation, reduce self-heating, and improve frequency stability. These results highlight the potential of AlN TE-mode bulk acoustic resonators as a thermally robust platform for compact and efficient piezoelectric power conversion.

Acknowledgement

The authors appreciate Dr. Sunil Bhave for helpful discussions. This work was supported by the Defense Advanced Research Projects Agency (DARPA) High Operational Temperature Sensors (HOTS) program and the DARPA Nimble Ultrafast Microsystems (NIMBUS) program. Any opinions, findings, conclusions, or recommendations expressed in this material are those of the author(s) and do not necessarily reflect the views of DARPA.

REFERENCES

[1] K. Nguyen *et al.*, "Near spurious-free thickness shear mode lithium niobate resonator for piezoelectric power conversion," *IEEE Trans. Ultrason., Ferroelectr., Freq. Control*, vol. 70, no. 11, pp. 1536–1543, 2023.

[2] D. T. Brown *et al.*, "A new class of topologies for isolated piezoelectric-based power conversion," in *Proc. IEEE 26th Workshop Control Model. Power Electron.* (COMPEL), 2025.

[3] Nguyen, K., *et al.*, "Near-spurious-free lithium niobate resonator for piezoelectric power conversion with $Q$ of 3500 and $k_t^2$ of 45%," in Proc. IEEE Int. Ultrasonics Symp. (IUS), 2022.

[4] Sullivan, C. R., *et al.*, "On size and magnetics: Why small efficient power inductors are rare," in Proc. 2016 Int. Symp. 3D Power Electron. Integr. Manuf. (3D-PEIM), 2016.

[5] J. D. Boles *et al.*, "Enumeration and analysis of DC–DC converter implementations based on piezoelectric resonators," *IEEE Trans. Power Electron.*, vol. 36, no. 8, pp. 9156–9170, Aug. 2021.

[6] S. Naval *et al.*, "High-efficiency isolated piezoelectric transformers for magnetic-less dc-dc power conversion," *IEEE Trans. Power Electron.*, 2026.

[7] T. J. Skinner, M. Touhami, and J. D. Boles, "A piezoelectric-resonator-based 'active inductor,'" in *Proc. IEEE Workshop Control Model. Power Electron.* (COMPEL), 2024.

[8] J. D. Boles *et al.*, "A piezoelectric-resonator-based DC–DC converter demonstrating 1 kW/cm resonator power density," *IEEE Trans. Power Electron.*, vol. 38, no. 3, pp. 2811–2815, 2022.

[9] W.-C. B. Liu, G. Pillonnet, and P. P. Mercier, "An integrated dual-side series/parallel piezoelectric resonator-based DC-DC converter," IEEE J. Solid-State Circuits, vol. 59, no. 12, pp. 4162–4174, Dec. 2024.

[10] E. A. Stolt, C. Daniel, and J. M. Rivas-Davila, "A stacked radial mode lithium niobate transformer for dc-dc conversion," in *Proc. IEEE Workshop Control Model. Power Electron.* (COMPEL), 2024.

[11] W. D. Braun *et al.*, "A stacked piezoelectric converter using a segmented IDT lithium niobate resonator," *IEEE Open J. Power Electron.*, vol. 5, pp. 286–294, 2024.

[12] C. Daniel *et al.*, "Nonlinear losses and material limits of piezoelectric resonators for DC–DC converters," in *Proc. IEEE Appl. Power Electron. Conf. Expo.* (APEC), 2024.

[13] A. Marques *et al.*, "First 6 MHz 8-phases regulated LNO based piezoelectric DC-DC converter," in Proc. 2025 IEEE 10th Southern Power Electron. Conf. (SPEC), 2025.

[14] M. Touhami *et al.*, "Piezoelectric materials for the DC-DC converters based on piezoelectric resonators," in Proc. 2021 IEEE 22nd Workshop Control Model. Power Electron. (COMPEL), 2021.

[15] E. Stolt *et al.*, "A spurious-free piezoelectric-resonator-based 3.2 kW DC–DC converter for EV on-board chargers," *IEEE Trans. Power Electron.*, 2024.

[16] Z. Yao *et al.*, "Lithium tantalate bulk acoustic resonator for piezoelectric power conversion," in *Proc. Int. Conf. Solid-State Sensors, Actuators, Microsyst.* (Transducers), 2025.

[17] Z. Cheng *et al.*, "Experimental observation of high intrinsic thermal conductivity of AlN," Phys. Rev. Mater., vol. 4, no. 4, Art. no. 044602, Apr. 2020.

[18] M. S. B. Hoque *et al.*, "High in-plane thermal conductivity of aluminum nitride thin films," ACS Nano, vol. 15, no. 6, pp. 9588–9599, Jun. 2021.

[19] G. A. Slack, "The intrinsic thermal conductivity of AlN," *J. Phys. Chem. Solids*, vol. 48, no. 7, 1987.

[20] A. V. Inyushkin *et al.*, "On the thermal conductivity of single crystal AlN," *J. Appl. Phys.*, vol. 127, no. 20, Art. no. 205109, 2020.

[21] X. Xiao *et al.*, "Performance of LiTaO3 crystals and thin films and their application," *Crystals*, vol. 13, no. 8, Art. no. 1233, 2023.

[22] U. Bashir *et al.*, "Thermal conductivity in solid solutions of lithium niobate tantalate single crystals from 300 K up to 1300 K," *J. Alloys Compd.*, vol. 1008, Art. no. 176549, 2024.

[23] G. A. Alvarez *et al.*, "Thermal conductivity enhancement of aluminum scandium nitride grown by molecular beam epitaxy," *Mater. Res. Lett.*, vol. 11, no. 12, pp. 1048–1054, 2023.

[24] Y. Song *et al.*, "Thermal conductivity of aluminum scandium nitride for 5G mobile applications and beyond," *ACS Appl. Mater. Interfaces*, vol. 13, no. 16, pp. 19031–19041, 2021.

[25] S. Yarlagadda *et al.*, "Low temperature thermal conductivity, heat capacity, and heat generation of PZT," *J. Intell. Mater. Syst. Struct.*, vol. 6, no. 6, pp. 757–764, 1995.

[26] R. L. Xu *et al.*, "Thermal conductivity of crystalline AlN and the influence of atomic-scale defects," *J. Appl. Phys.*, vol. 126, no. 18, 2019.

[27] M. S. B. Hoque *et al.*, "High in-plane thermal conductivity of aluminum nitride thin films," *ACS Nano*, vol. 15, no. 6, pp. 9588–9599, 2021.

[28] Y. Fu *et al.*, "Origin of the difference in thermal conductivity and anharmonic phonon scattering between $LiNbO_3$ and $LiTaO_3$," *CrystEngComm*, vol. 23, no. 48, pp. 8572–8578, 2021.

[29] Z. Ju *et al.*, "Lattice thermal conductivity of $LiNbO_3$ considering high-order anharmonicity and phonon mutual coherence channel," *Phys. Rev. Appl.*, vol. 24, no. 2, Art. no. 024035, 2025.

[30] M. Tachibana, T. Kolodiazhnyi, and E. Takayama-Muromachi, "Thermal conductivity of perovskite ferroelectrics," *Appl. Phys. Lett.*, vol. 93, no. 9, Art. no. 092902, 2008.

[31] B. M. Foley *et al.*, "Phonon scattering mechanisms dictating the thermal conductivity of lead zirconate titanate ($PbZr_{1-x} Ti_x O_3$) thin films across the compositional phase diagram," *J. Appl. Phys.*, vol. 121, no. 20, Art. no. 205104, 2017.

[32] E. Langenberg *et al.*, "Ferroelectric domain walls in $PbTiO_3$ are effective regulators of heat flow at room temperature," *Nano Lett.*, vol. 19, no. 11, pp. 7901–7907, 2019.

[33] D. Bugallo *et al.*, "Deconvolution of phonon scattering by ferroelectric domain walls and point defects in a $PbTiO_3$ thin film deposited in a composition-spread geometry," *ACS Appl. Mater. Interfaces*, vol. 13, no. 38, pp. 45679–45685, 2021.

[34] K. Nguyen *et al.*, "Spurious-free lithium niobate bulk acoustic resonator for piezoelectric power conversion," in *Proc. Joint Eur. Freq. Time Forum IEEE Int. Freq. Control Symp.* (EFTF/IFCS), 2023.

[35] V. Chulukhadze *et al.*, "Lithium niobate resonators for power conversion: Spurious mode suppression via an active ring," in *Proc. IEEE Ultrasonics, Ferroelectrics, and Frequency Control Joint Symp.* (UFFC-JS), 2024.

[36] Z. Yao *et al.*, "Periodically poled piezoelectric lithium niobate resonator for piezoelectric power conversion," *arXiv* preprint arXiv:2508.09407, 2025.